\documentstyle[12pt,cite,epsfig]{article}

\newcommand{\coef}
{\gamma \left(2\omega_{p}{\frac{\hbar}{m}}\right)^{-1/2}}

\newcommand{\Journal}[2]{#1, #2}
\newcommand{\Book}[5]{#1, {\it #2} #3, #4, (#5)}
\newcommand{\Author}[2]{#1. #2}
\newcommand{\Title}[5]{{\it #1} #2 {\bf #3}, #4, (#5)}

\newcommand{\TMF}{Teor.Mat.Fiz.}
\newcommand{\TMP}{Theor.Math.Phys.}
\newcommand{\ZETF}{Zh.Eksp.Teor.Phys.}
\newcommand{\AMEP}{Appl.Math.\&Eng.Phys.}

\newcommand{\QM}{Quantum Mechanics - Non-relativistic Theory}
\newcommand{\PNF}{Physical Nature of the Fireball}
\newcommand{\SPSM}{Some Problems of Statistical Mechanics}
\newcommand{\PK}{Physical Kinetics}
\newcommand{\FT}{The Classical Theory of Fields}

\begin{document}

\centerline{{\Large {\bf Approximate Solutions of Quantum Equations}}}
\centerline{{\Large {\bf for Electron Gas in Plasma}}}

\centerline{{\large {\bf M.Dvornikov
\footnote{e-mail: maxim\_dvornikov@aport.ru}}}}
\centerline{\bf Department of Theoretical Physics,
Moscow State University}
\centerline{\bf 119992 Moscow, Russia}

\centerline{{\large {\bf S.Dvornikov}}}
\centerline{\bf Acoustical Institute, 'Shvernika' street, 4,}
\centerline{\bf 117036 Moscow, Russia}

\centerline{{\large {\bf G.Smirnov}}}
\centerline{\bf Kamchatka Hydrophysical Institute,}
\centerline{\bf KHPI, Kamchatka region, Viliutchinsk, Russia}

\centerline{{\it Abstract}}

{\it
We have obtained the  solutions of linearized Shr{\"o}dinger equation
for spherically and axially symmetrical electrons density oscillations
in plasma in the approximation of the self-consistent field.
It was shown that in the center or on the axis of symmetry of
such a system  the static density of electrons can enhance,
which leads to the increasing of density and pressure
of ion gas. We suggest that this mechanism could be
realized in nature as rare phenomenon
called the 'fireball' and could be used in
carrying out the research concerning controlled fusion.}

If a volume charge appears in electroneutral plasma, i.e.
electrons density increases or decreases in some finite area, then,
after an external influence is over, the oscillating process
consisting in periodical changes of the sign of the considered
volume charge is known to appear. It is necessary to remind that
the frequency of this process $\omega_{p}$ called plasma frequency is
related to free electrons density in the plasma $n_{0}$ by the
formula
\begin{equation}
\omega _{p}^{2} =
\frac{{4 \pi e^{2}n_{0}}}{{m}},
\end{equation}
where $e$ and $m$ are the charge and the mass of the electron.

We will assume electrons
in plasma to be a quantum many body system. Such an
assumption is due to the fact that, as it
will be shown below, the solutions obtained have
characteristic sizes of atomic order. It is known that the classical
dynamics of $N$ interacting particles can be represented by the system
of differential equations of motion in configuration space of
$3N$-dimensions,
or in $6N$-dimensional phase space, as well as by the
system of partial differential equations in three-dimensional
physical space and as the dynamics of singular material fields
(see Refs.~\cite{La.Li/FT(88),La.Li/QM(89),Dr.Ku/TMP(96)}).
The transfer to the quantum-mechanical description is realized
by changing the dynamic functions to Hermitian operators. In this
case the dimensionality of the configuration space is conserved and
the state of the system is completely defined by the wave functions
in the $3N$-dimensional space.

Shr{\"o}dinger equation as it was shown in
Refs.~\cite{Dr.Ku/TMP(96),Ma.Ku/TMP(99)},
could be presented in the form of the system of equations in
physical space and having the same form as the equations of
hydrodynamics. In these works the quantum-mechanical
system of $N$ particles with arbitrary masses and charges,
interacting between itself by Coulomb forces and with
external classical electromagnetic field characterized by vector
$\vec A$ and scalar $\varphi$ potentials has been considered.

In this approach the complex $\Psi$ function introduced in
three-dimensional
space has the following form
\begin{equation}
\Psi \left( {{\vec r},t} \right) =
\sqrt {n_{e} \left( {{\vec r},t} \right)}
e^{\frac{{i}}{{\hbar} }\sigma  \left({\vec r},t \right)},
\end{equation}
where $\left| \Psi \right|^{2}=n_{e}({\vec r},t)$
is the density of
electrons and $\Psi$ function satisfies the
partial differential equation:
\begin{equation}
i\hbar \frac{{\partial \Psi
}}{{\partial  t}} =
{\hat H} \Psi .
\end{equation}
The Hamiltonian in  Eq.~(3) is expressed in the following way
\begin{equation}
{\hat H} =
\frac{{1}}{{2m}}\left( {\frac{{\hbar} }{{i}}{\vec \nabla} -
\frac{{e}}{{c}}
{\vec A}\left( {\vec r},t \right)} \right)^{2} + e\varphi+
e^{2}\int\limits_{{\vec r}_{0}} ^{\vec r} {d^{3}{\bf r}^{\prime}
G\left( {\vec r}-{\vec r}^{\prime} \right)
\left| {\Psi}  \right|^{2} + \theta \left( {\vec r},t \right)} ,
\end{equation}
where $G\left( {\vec r}-{\vec r}^{\prime} \right)=
{\frac{1}{\left| {\vec r}-{\vec r}^{\prime} \right|}}$.

In the Eq.~(4) the first two terms are the components of
one electron Hamiltonian in the external
electromagnetic field, the third term represents the potential energy
of the electron in the self-consistent electrostatic field
created by the whole electron system, with density
of number of particles being equal to
$\left| \Psi \right|^{2}$. The function
$\theta \left( {\vec r},t \right)$
describing exchange interactions between electrons has the
following form
\begin{equation}
\theta  \left( {\vec r},t \right)=
\frac{{\hbar ^{2}}}{{2m}}\frac{{\Delta \Psi} }{{\left| {\Psi}
\right|}} + \int\limits_{ {\vec r}_{0}}^{\vec r}
{\frac{{dp_{e} \left(
{\vec r},t \right)}}{{\left| {\Psi}  \right|^{2}}}} + e^{2}\int
{d^{3}{\bf r}^{\prime}\int\limits_{{\vec r}_{0}}^{\vec r}
{d}G\left( {{\vec r} - {\vec r}^{\prime}} \right)
\frac{{q_{2} \left( {{\vec r},{\vec r}^{\prime},t} \right)}}{{\left|
{\Psi}  \right|^{2}}}} ,
\end{equation}
where $p_{e} \left( {\vec r},t \right)$
is the pressure of electron
gas, $q_{2} \left( {\vec r},{\vec r}^{\prime},t \right)$ is the
correlation function.

Therefore, in order to resolve exactly the considered problem one
should take into account all terms in the Eq.~(5). However, we will
assume that the contribution of exchange
interactions to the dynamics
of free electrons in plasma is much smaller than that of the
potential of self-consistent electrostatic field. Then, in
describing the dynamics of electron gas we will neglect
the function
$\theta \left( {\vec r},t \right)$. As it will be
seen from further speculations, this rough approximation allows
us to get some consequences which are close to those
obtained from the consideration of the similar problem within the
classical approach.

Let us consider the neutral plasma, formed
by the singly ionized gas, with the energy of electrons being more
than the potential of ionization of this gas. We will suppose
that this plasma possesses a spherical symmetry for density and
velocities distribution of the electron and the ion gases.
Taking into account the small mobility
of heavy ions in gas
compared to that of electrons we will suppose that the density of
ions is the constant value
$n_{i} \left( {\vec r},t \right) = n_{0}$.
We will also consider that in our case there are no external
electromagnetic fields except those of positively charged ions.
It is worth mentioning that the magnetic field in the
system is equal to zero. The potential of self-consistent field
created by the density of electron gas is represented by the
formula (using spherical coordinates):
\begin{equation}
U_{e}=e\int\limits_{{\vec r}_{0}}^{\vec r}
d^{3} {\bf r} G({\vec r}^{\prime}-{\vec r})
{{\left|{\Psi}  \right|^{2}}} =
4{\pi}e\int\limits_{r}^{\infty} {\frac{dR}{R^{2}}}
\int\limits_{0}^{R} x^{2} {{\left|{\Psi}(x,t) \right|^{2}}}dx.
\end{equation}
Similarly,
for the potential $\varphi$ of singly ionized gas with
density of ions $n_{i}(r,t)$ one has
\begin{equation}
\varphi =-4{\pi}e\int\limits_{r}^{\infty} {\frac{dR}{R^{2}}}
\int\limits_{0}^{R} x^{2} n_{i}(x,t) dx.
\end{equation}
Thus, taking into account
Eq.~(6) and Eq.~(7), the Eq.~(3) can be represented in the
following way
\begin{equation}
i\hbar \frac{{\partial \Psi} }{{\partial
t}} + \frac{{\hbar ^{2}}}{{2m}}\Delta
\Psi - 4\pi  e^{2}
\Psi F({{\left|{\Psi} \right|^{2}}}-n_{0})=0,
\end{equation}
where
$\Delta={\frac{\partial^{2}}{\partial r^{2}}}+
{\frac{2}{r}}{\frac{\partial}{\partial r}}$ is the Laplas operator
in the spherical coordinate system,
$F(\ldots)=\int\limits_{r}^{\infty} {\frac{dR}{R^{2}}}
\int\limits_{0}^{R} x^{2} (\ldots) dx.$

Moreover, we demand the system to be electroneutral as a whole,
i.e. the condition must be satisfied:
\begin{equation}
\lim_{R\rightarrow\infty} {\frac{1}{R^{3}}}
\int\limits_{0}^{R} x^{2} {{\left|{\Psi}(x,t) \right|^{2}}}dx=
{\frac{n_{0}}{3}}.
\end{equation}

We will search for a solution of the Eq.~(8) in the form:
\begin{equation}
\Psi = \Psi _{0} + \chi e^{ - i\omega t} \quad .
\end{equation}

Supposing that $\left| \Psi _{0} \right|^{2}=n_{0}$ and
$\left| \chi \right| \ll \left| \Psi_{o} \right|$, we get
$\left| \Psi_{0} \right|^{2}=
n_{0} + \Psi_{0} f +\left| \chi \right|^{2}$, where
$f=\chi e^{-i\omega t}+\chi^{*} e^{i\omega t}$. Taking into
account that ${\left| \chi \right|} \ll \left| \Psi_{o} \right|$, we
obtain $\left| \Psi_{0} \right| ^{2}\approx n_{0}+\Psi_{0} f$.

Then, we substitute Eq.~(10) and approximate expression for
$\left| \Psi_{0} \right|^{2}$ in Eq.~(8). Having considered the
complex-conjugate
equation together with the obtained one, it was easy to get:
\begin{equation}
\hbar \omega  f + \frac{{\hbar ^{2}}}{{2m}}\Delta f
- 4\pi  e^{2}\Psi _{0}
\left( {2\Psi _{0} + f} \right) F\left( {f} \right)=0.
\end{equation}

For the total linearization, it is necessary to suppose that
$2\Psi _{0} + f \cong 2\Psi _{0}$ in the third term of the
equation involved. Then, we represent the function $\chi$ through its
real and imaginary parts: $\chi=\chi_{1}+i\chi_{2}$. Hence
$f=\chi_{1}\cos\omega t+\chi_{2}\sin\omega t$,
and Eq.~(8) can be divided
into two independent similar equations for $\chi_{1}$
and $\chi_{2}$:
\begin{equation}
\hbar \omega  \chi _{n} + \frac{{\hbar
^{2}}}{{2m}}\Delta  \chi _{n} - 8\pi
e^{2}n_{0}  F\left( {\chi_{n}}  \right)=0, \quad n = 1,2.
\end{equation}

\begin{figure}[htb]
\begin{center}
\epsfig{file=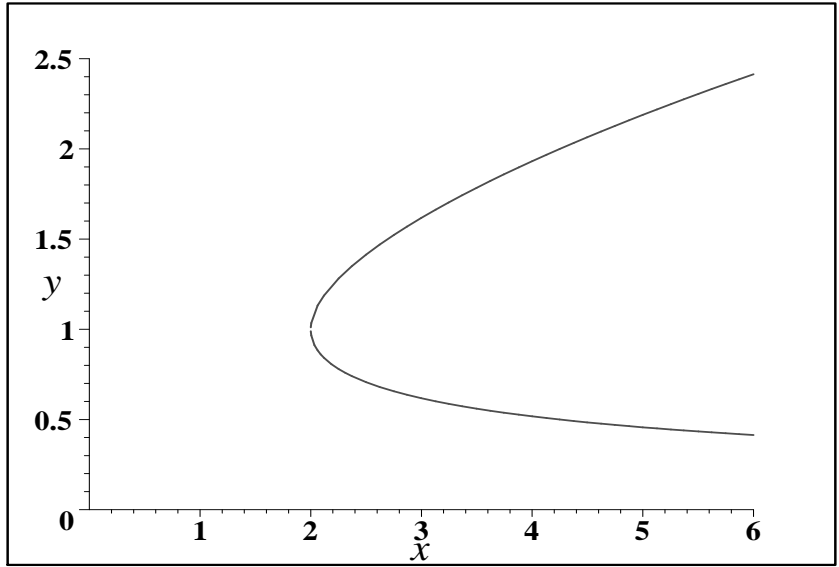,height=7.5cm,width=11.0cm}
\end{center}
\caption{\label{disprel} The coeficient
$y=\coef$
versus the parameter
$x={\frac{\omega}{\omega_{p}}}$.
If
$\omega=2\omega_{p}$, then
$\gamma_{1,2}= \pm \left( {\frac{2m\omega_{p}}{\hbar}}
\right)^{1/2}$.
If
$\omega \gg 2\omega_{p}$,
for the upper branch we
have
$\gamma_{1,2}= \pm \left( {\frac{2m\omega}{\hbar}} \right)^{1/2}$,
and for the lower branch
$\gamma_{3,4}= \pm \left(
{\frac{2m\omega_{p}^{2}}{\omega\hbar}} \right)^{1/2}$.
}
\end{figure}

One can find out the functions
$\chi _{n} = B_{n}\frac{{sin\gamma r}}{{r}}$ are the solutions of of
the Eq.~(12) if $\gamma$ satisfies the following
dispersion relation
\begin{equation}
\gamma^{2}={\frac{\omega m}{\hbar}}
\left[
1\pm
\left(
1-4{\frac{\omega_{p}^{2}}{\omega^{2}}}
\right)^{1/2}
\right],
\end{equation}
where $\omega_{p}$ is determined in Eq.~(1). The
positive part of the dependence of
$\gamma$
on the frequency $\omega$
is presented in the Fig.~\ref{disprel}. It is worth noticing that if
the values of $B_{n}$ are limited, the expressions for $\chi_{n}$
satisfies the condition of system electroneutrality (9).

From the expression (13) one can see that the frequency
$\omega=2\omega_{p}$ is the critical value, since for frequencies
less than $2\omega_{p}$, $\gamma$ becomes a complex value and under
these circumstances the oscillations of the electron gas are absent.
It is necessary to remind that in using the classical approach to the
similar problem one gets the value $\omega=\omega_{p}$ for the
critical frequency.

For example, for the density $n_{0}=2.7\cdot 10^{19} cm^{-3}$, i.e.
for completely singly ionized gas under the atmospheric pressure
and when $\omega=2\omega_{p}$, the frequency of electron oscillations
is:
$$
\nu={\frac{\omega_{p}}{\pi}}=
2\left( {\frac{e^{2}n_{0}}{\pi m}} \right)^{1/2} \approx
9\cdot 10^{13} Hz.
$$

This frequency corresponds to the electromagnetic radiation in the
infrared range with the wavelength
$\lambda={\frac{c}{\nu}}\approx 3\cdot 10^{-4} cm$. In this case
$\gamma_{1,2}=
\pm \left( {\frac{2m\omega_{p}}{\hbar}} \right)^{1/2}
\approx \pm 2.3\cdot 10^{7} cm^{-1}$, the size of
central region $\delta$,
where the most intensive oscillations of the
electron gas are observed, is equal to
$\pi / \gamma \approx 1.4\cdot 10^{-7} cm$.

It is necessary to remind that the exact expression for
the density of electron gas in searching for the solution in
the form of the Eq.~(10) is presented as
$\left| \Psi_{0} \right|^{2}=
n_{0} + \Psi_{0} f +\left| \chi \right|^{2}$. Let us define in this
expression $\Psi_{0} f$ and
${\bar n}_{e}=\left| \chi \right|^{2}+n_{0}$ as the dynamic and
static components of density respectively. In deriving the
approximate linearized equation (12) the function
$\left| \chi \right|^{2}$ was supposed to be small
and thus neglected.
This procedure is not correct because the integral
$F\left( {\left| \chi \right|^{2}} \right)$ for the function
$\chi = B\frac{{sin\gamma r}}{{r}}$ is divergent.  However, we
assumed that the density of ion gas was constant throughout the
volume because of the small mobility of heavy ions.
This assumption is valid
for the frequently oscillating dynamic component, but
$\left| \chi \right|^{2}$ does not depend
on time. Therefore, it is naturally to expect that under some
conditions the negative volume charge, described by this
function, will be compensated (or neutralized) by removing
positive ions, that will result in local changing of density and
pressure of the ion gas.

Indeed, it can be shown that for our case the condition of
static stability of the ion gas is
(see, for instance, Refs.~\cite{Bo.Sa/SPSM(75),Li.Pi/PK(79)}):
\begin{equation}
kT
{\frac{\partial n_{i}}{\partial r}}=
{\frac{4\pi e^{2}}{r^{2}}}
n_{i}
\int\limits_{0}^{r}
\left( n_{i} - {\bar n}_{e} \right) x^{2}dx
\end{equation}
where $k$ is Boltsman constant, $T$ is the ion gas temperature,
$n_{i}$ is the ions density, ${\bar n}_{e}$ is the static electrons
density.

We demand that the ions density is equal to static electrons
density with high level of accuracy
$n_{i}\approx {\bar n}_{e}=
n_{0}+B^{2}{\frac{\sin^{2}\gamma r}{r^{2}}}$. Then, from Eq.~(14) we
get the following inequality:
${\frac{kT}{4\pi e^{2}}}{\frac{r^{2}}{n_{i}}}
\left|{\frac{\partial n_{i}}{\partial r}}\right|\ll 1$.
Having substituted the expression for
$n_{i}$ in the last formula,
we obtained the condition of the neutralization:
${\frac{kTB^{2}\gamma}{4\pi e^{2}n_{0}}}\ll 1$. Taking into account
that ions density in the center of the system should be equal to
$n_{c}=B^{2}\gamma^{2}$, this condition can be rewritten in the
following way:
\begin{equation}
n_{c}\ll {\frac{4\pi e^{2}\gamma}{kT}}n_{0}.
\end{equation}
For instance, for $T=1000^{0} K$ and the value
$\gamma=2.3\cdot 10^{7} cm^{-1}$
which was obtained above we get $n_{c}\ll 500 n_{0}$. In deriving the
solution of our problem we assumed that $n_{c}\ll n_{0}$. Thus, we can
conclude that, if the condition (15) is satisfied, the ion charge
is undoubtedly compensated by the supplementary static
electron charge and the divergence in the integral
$F(|\Psi|^{2}-n_{i})$
can be eliminated.

Hence, the approximate, linearized theory
trends to describe the pressure enhancement
of the ion gas in the center of symmetrically oscillating electron
gas.

It is worth mentioning that along with spherically symmetrical
solution of the Eq.~(12), there is at least one axially-symmetrical
one which has the form:
$$\Psi = n_{0}^{1/2}+B J_{0}(\gamma r)
e^{-i\omega t},$$
where $J_{0}$ is the zero-order Bessel function.
In this case the dispersion relation takes the same form as Eq.~(13).
All consequences obtained for spherically-symmetrical oscillations
are valid and for this case too.

While considering nonlinear
Shr{\"o}dinger equation (8), it can be seen that along with the
components of frequency $\omega$, the terms which do not depend
on time as well as on frequencies $2\omega$, $3\omega$ and etc.
appear.
One can make sure of this representing
the solution of the Eq.~(8) in the form:
$\Psi (r,t)=\Psi_{0}+\sum\limits_{k=1}^{\infty} \Psi_{k}(r,t),$ where
$\Psi_{1}=B_{1}e^{-i\omega t}{\frac{\sin \gamma r}{r}}$.

In Ref.~\cite{Dv.Dv.Sm/AMEP(01)} we put forward the hypothesis that
in this case the static density of electron gas could attain great
values. Density and
pressure of ion gas are also great in the center of such systems.
Thus, along with the heating of the central area resulted from
intensive movement of electrons, there will be a possible running of
nuclear fusion reactions of corresponding nuclei in gas.

This process seems to support the
existence of the enigmatic natural phenomena, known as
'fireball'. If vapors of water which are present in atmosphere
contain deuterium in the amount of $\approx 1/5000$ under normal
conditions, the running the nuclear fusion reactions will release
the energy, which support the oscillations of electron gas and
prevent the recombination of plasma. Taking into account small sizes
of the central (active) regions and small amount of
deuterium in atmosphere, it is possible to
use the
term 'microdose' thermonuclear reaction for process in question.
Axially-symmetrical
oscillation of the electron gas is likely to appear as very seldom
observed type of a 'fireball' in the form of shining, sometimes
closed cord \cite{St/PNF(85)}.
Uncomplicated calculation shows that energy released
in deuterium nuclei fusion in $1 dm^{3}$ of vapors of water (the
average size of a 'fireball') has the value of about $1 MJ$, that
corresponds to evaluations of energy released by some observed
'fireballs' (see, for example, Refs.~\cite{St/PNF(85),Fr/ZETF(40)}).

Groups and separate researchers developing the problem of
the controlled fusion are suggested to pay attention
to self-consistent, radially and axially oscillating
electron plasmoids as a base models to
self-supported thermonuclear reactions. The authors of the present
article have certain experience in generating
ball-like plasma structures which proves the correctness of the
chosen model.

\paragraph{Acknowledgements}

The authors are indebted to the members of the seminar on Theoretical
Physics (General Physics Institute of RAS, Moscow, October 1999) for
their helpful comments which were taken into account in preparing of
the present paper.

\end{document}